\documentclass[aps,pra,showpacs,showkeys,twocolumn,superscriptaddress]{revtex4-1}

\usepackage{graphicx,tabularx,longtable,color,enumitem,times,hyperref}

\setcitestyle{numbers,square}

\begin{document}

\title{Exploring student difficulties with observation location}
\author{Jaime Bryant}
\email{jbry92@gmail.com}
\affiliation{DePaul University, College of Education, 2247 N. Halsted St, Chicago, IL 60614, USA}
\author{Rita Dawod}
\author{Susan M. Fischer}
\author{Mary Bridget Kustusch}
\affiliation{DePaul University, Department of Physics, 2219 North Kenmore Avenue Suite 211, Chicago, IL 60614, USA}

\date{\today}

\begin{abstract}
\noindent\textbf{Abstract.}
Throughout introductory physics, students create and interpret free body diagrams in which multiple forces act on an object, typically at a single location (the object's center of mass).  The situation increases in difficulty when multiple objects are involved, and further when electric and magnetic fields are present. In the latter, sources of the fields are often identified as a set of electric charges or current-carrying wires, and students are asked to determine the electric or magnetic field at a separate location defined as the observation location. Previous research suggests students do not always appropriately account for how a measurement or calculation depends on the observation location. We present preliminary results from a studio-style, algebra-based, introductory electricity and magnetism course showing the prevalence of correct and incorrect responses to questions about observation location by analyzing student written work involving vector addition of fields.
\end{abstract}


\maketitle

\section{introduction\label{intro}}
A significant portion of introductory electricity and magnetism courses requires students to identify the electric or magnetic field at a specific location due to one or more sources. 
The position of this `observation location' relative to the source (\emph{i.e.,} the separation vector, $\vec{r}-\vec{r}'$) is one of the most important quantities for determining the field at that location. 
When we represent a vector field in space, we place the tail of each vector at a different observation location. These vectors will differ in direction and magnitude based on the separation vector  between the observation location and the source(s). 

In this paper, we present a first piece of a larger study exploring student understanding of the spatial nature of electric and magnetic fields, for which the data are approximately 180 hours of classroom video and all student written and online work from three sections of a studio-style, algebra-based, introductory electricity and magnetism course. 
Given previous research showing that students often do not appropriately account for the observation location~\cite{Scaife:2011,Kustusch:2015}, 
we analyzed student responses to quiz questions where they were asked to draw the electric or magnetic field at an observation location due to multiple sources.
The observation location impacts where the vector should be placed, what direction it will point, and how large it will be. Thus, we examined all three of these aspects of student responses as a window into how students are (or are not) accounting for the observation location.


\section{Methods\label{methods}}
\begin{figure}[b]
\fbox{
\begin{minipage}{3.3in}
\smallskip
\begin{description}
\item[Problem] A positive and negative charge of equal but opposite charge are located as shown in the figure below.  The negative charge is twice as far away from the asterisk as the positive charge.  With reasonable accuracy, draw a vector to indicate the direction of the net electric field at the asterisk in the figure.  You may redraw the figure if you like.  If you draw more than one vector, please clearly indicate the vector that represents the net electric field. 
\begin{center}
\includegraphics[scale=0.35]{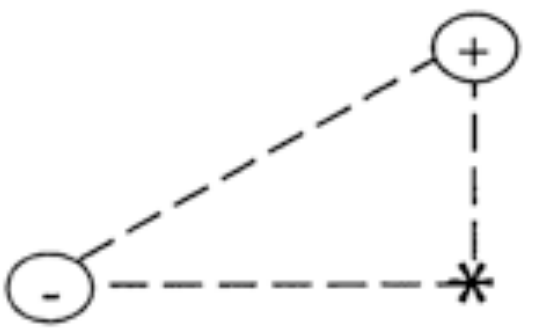}
\end{center}
\end{description}
\caption{Question for Quiz 1 used for Section A}
\label{quiz2a}
\end{minipage}
}
\end{figure}

\begin{figure}
\fbox{
\begin{minipage}{3.3in}
\smallskip
\begin{description}
\item[Problem] Two identical charges are located as shown in the figure below. The top charge is twice as far away from the asterisk as the lower charge. With reasonable accuracy, draw a vector to indicate the direction of the net electric field at the asterisk in the figure below. You may redraw the figure if you like. If you draw more than one vector, please clearly indicate the vector that represents the net electric field.
\end{description}
\includegraphics[scale=0.55]{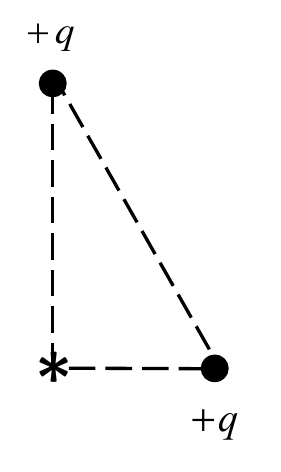}
\caption{Question for Quiz 1 used for Sections B and C.}
\label{quiz2bc}
\end{minipage}
}
\end{figure}

\begin{figure}
\fbox{
\begin{minipage}{3.3in}
\smallskip
\begin{description}
\item[Problem] Two parallel, straight wires each carry a current of I, and both currents (and wires) are directed out of the page. With reasonable accuracy, i.e. approximately to scale, and in the coordinate system shown below, draw a graphical representation of the vector addition used to find the net magnetic field at the asterisk ($^*$). Please label your vectors.
\end{description}
\includegraphics[scale=0.55]{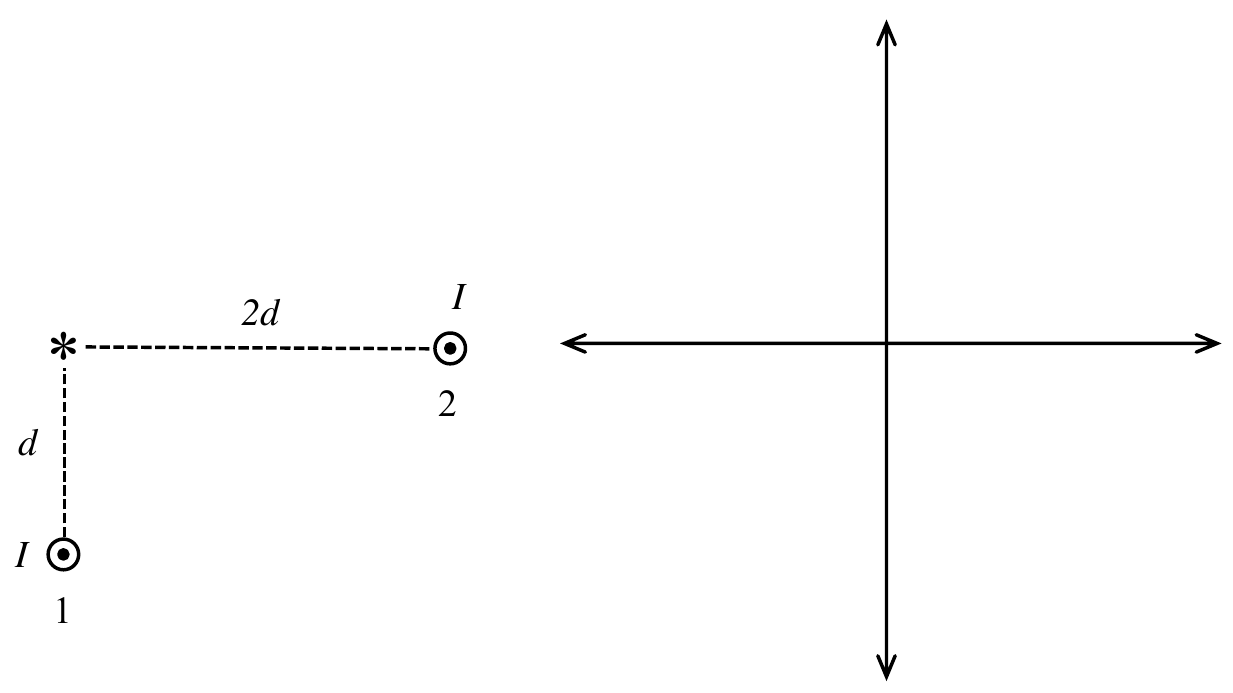}
\caption{Question from Quiz 2, which was used for all sections.}
\label{quiz6}
\end{minipage}
}
\end{figure}

Our goal with this initial study was to explore how this algebra-based population would respond to questions about observation location after studio-style instruction, with the intention of using these results to guide a deeper qualitative analysis using classroom video.
%
We analyzed  written responses to select questions from two quizzes on electric and magnetic fields (Figs.~\ref{quiz2a}--\ref{quiz6}). 
Quiz~1 was administered during the second week of a ten-week quarter and each section included a question that assessed superposition of electric fields from source charges (see Figs.~\ref{quiz2a}--\ref{quiz2bc}). 
Quiz~2 was given during the seventh week of the quarter and all sections used the same question to assess superposition of magnetic fields from source currents (see Fig.~\ref{quiz6}) . 
In each question, students were asked to draw the net electric (or magnetic) field due to two different sources at an asterisk identifing the observation location. 
For the Quiz~2 question, the instructors chose to provide a separate coordinate system where the student was asked to represent the vector addition used to determine the net magnetic field.
These questions  were chosen for their potential to highlight student understanding of observation location.

Since our focus is not on vector addition, we analyzed vectors representing the fields from each source and not the resultant vector (unless only a resultant vector was drawn). 
We initially focused on identifying where students placed the vectors.
Then, for those students who appeared to place the vectors at the given observation location, we determined if the direction of the vectors was correct and if so, whether the relative magnitude of the vectors was correct.  

\section{Vector Placement}\label{vector placement}
In each of the questions, the observation location was explicitly noted with an asterisk, so we first sought to determine whether students recognized that the field vectors should actually be placed at the asterisk. Using an iterative and emergent coding process with multiple coders, we identified three different vector placements based on the data, defined in Table~\ref{placement}.

\begin{table}
\caption{Categories for identifying where students drew arrows representing the electric or magnetic fields/forces.}
\label{placement}
\begin{ruledtabular}
\begin{tabular}{p{0.5in}p{2.75in}} 
Name& Category Description\\
\hline
Source& Student places at least one arrow with the tail of the arrow at a point charge or current.\\
\\
Asterisk&Student places the tail of the arrow(s) at the asterisk on the diagram and/or on a separate coordinate system where the asterisk is explicitly identified.\\
\\
Without Asterisk & Student only draws arrows on a separate coordinate system without explicitly identifying the asterisk.\\
\\
Unclear&Student places arrows at some other location and/or solution was unclear such that it could not be categorized.\\
 \end{tabular}
\end{ruledtabular}
\end{table}

\begin{table}
\caption{Results from both quizzes on where students placed arrows representing the electric or magnetic field.}
\label{placement results}
\begin{ruledtabular}
\begin{tabular}{lcc} 
& Quiz 1  &Quiz 2\\
&($N=95$)&($N=91$)\\
\hline
Source&13 (14\%)&17 (19\%)\\
Asterisk&70 (74\%)&40 (44\%)\\
Without Asterisk&6 (6\%)&30 (33\%)\\
Unclear&6 (6\%)&4 (4\%)
 \end{tabular}
\end{ruledtabular}
\end{table}

Other than the asterisk (`Asterisk'), students most often drew vectors  at one or both of the source(s) (`Source') and/or on a separate coordinate system drawn by the student (Quiz~1) or on the one provided (Quiz~2) (`Without Asterisk'). The number of responses in each category are shown in Table~\ref{placement results}.


\begin{figure}
\includegraphics[width=3.4in]{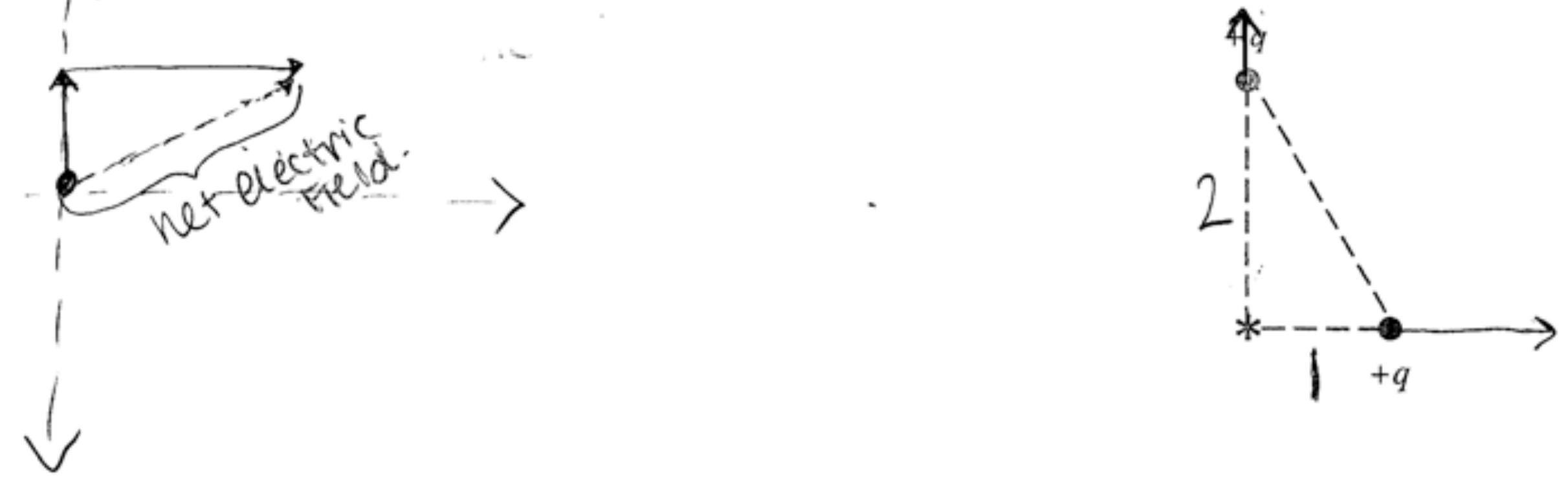}
\caption{Hope's solution to the question in Fig.~\ref{quiz2bc}, coded as `Source.'}
\label{Hope}
\end{figure}

 If any vector was drawn at a source, it was placed in the `Source' category. Approximately $15\%$ of the students on each quiz placed at least one arrow at a source. The solution from ``Hope'' (Fig.~\ref{Hope}) shows a typical example of this category from Quiz 1 (even though she includes a separate coordinate system, vectors at the sources take precedence in our coding).
 
One interpretation of Hope's solution is that she drew vectors that represent the direction of the electric force on the source charges from a positive test charge at the asterisk. This suggests confusion between field and force, consistent with previous research~\cite{Garza:2012,Kustusch:2015} or between source and test charges.
Supporting this interpretation is that one of the instructors informed us that during Quiz~1, several students would point to the asterisk and ask ``Is this a positive charge?''

For those students who used a separate coordinate system, some explicitly identified that the origin was the asterisk and/or also drew the vectors at the asterisk. These students were placed in the `Asterisk' group (\emph{e.g.,} Fig.~\ref{Gilma}) in contrast to the `Without Asterisk' group, who only drew vectors on a separate coordinate system without identifying that the net vector belongs at the asterisk (\emph{e.g.,} Fig.~\ref{Leslie}).

On the electric field questions, most students ($>70\%$) at least recognize that the vectors should be placed at the identified observation location. While this number drops dramatically for the magnetic field question, the difference in question format between Quiz 1 and Quiz 2 (\emph{i.e.}, the inclusion of a separate coordinate system on Quiz 2) makes it unclear how much of this difference is actually a reflection of where students think the vectors should be placed.

\section{Direction and Magnitude}\label{direction and magnitude}
\begin{figure}
\includegraphics[width=3.25in]{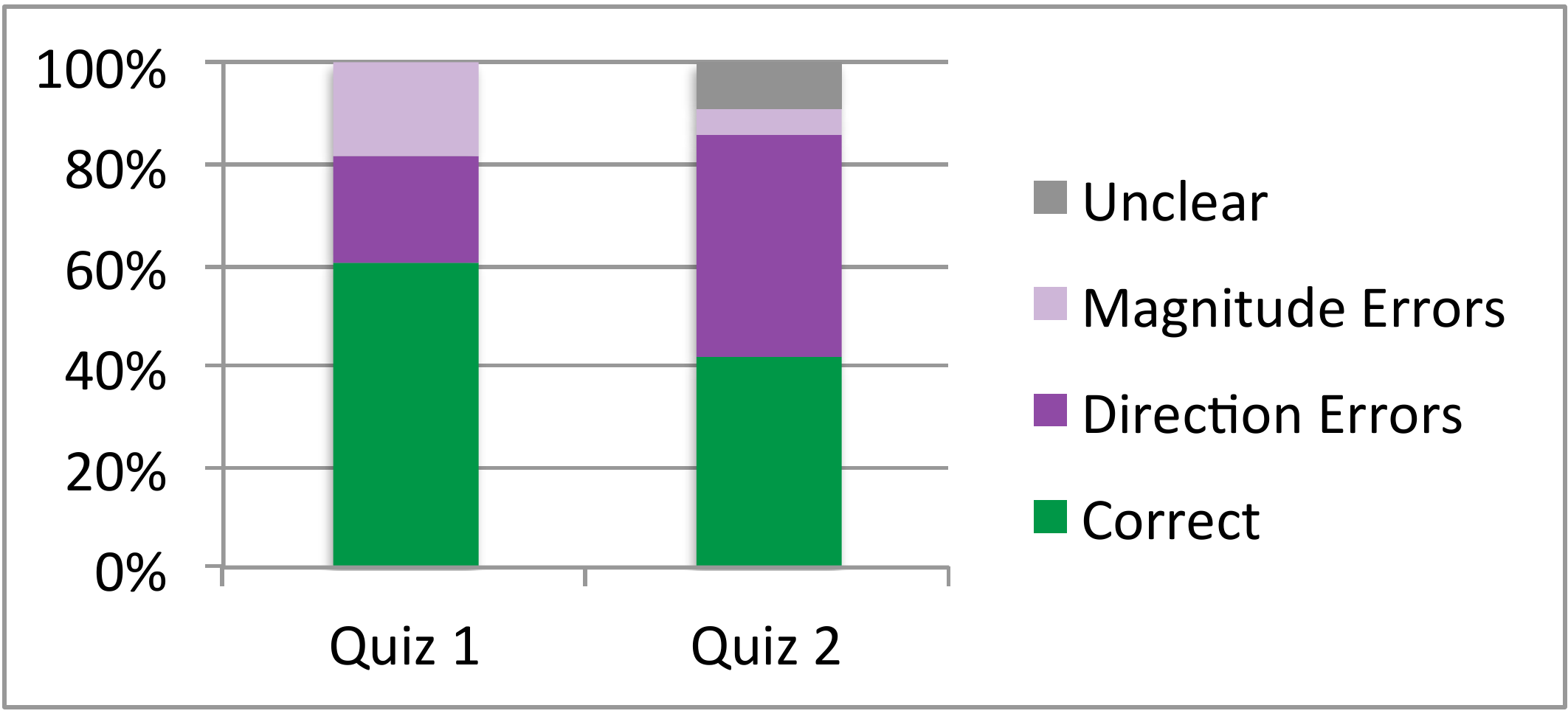}
\caption{Percentage of students on each quiz coded for correctness.}
\label{correctness}
\end{figure}

We examined answers of those in the `Asterisk' and `Without Asterisk' group for whether they correctly identified the direction and magnitude of the fields from each source based on their relative locations to the asterisk. 
If the direction of at least one vector was incorrect, it was coded as a `Direction Error', but if the directions were correct and the relative magnitudes were not, it was coded as a `Magnitude Error'. Since the relationship between the field and distance was different for the two questions (inverse square for Quiz 1 and inverse for Quiz 2), we only looked at whether the vector from the closer source was larger than the vector from the farther source when assessing magnitude. 
If the direction and relative magnitude were correct for the field from each source, the response was coded as `Correct', regardless of the resultant vector. Results from this coding are shown in Table~\ref{correctness results} and Fig.~\ref{correctness}.

\begin{table}
\caption{Results from both quizzes on the correctness of vectors representing the electric or magnetic field.}
\label{correctness results}
\begin{ruledtabular}
\begin{tabular}{lccc} 
&Quiz 1& Quiz 2\\
&($N=76$)&($N=70$)\\
\hline
Correct&46 (61\%)&29 (41\%)\\
Direction Error&16 (21\%)&31 (44\%)\\
Magnitude Error&14 (18\%)&4 (6\%)\\
Unclear&0 *0\%)&6 (9\%)
 \end{tabular}
\end{ruledtabular}
\end{table}

Comparing correctness on Quiz 1 and Quiz 2, we see that the number of students coded as correct decreases by about $20\%$ and the percentage of students who made direction errors is more than double in the magnetic field context than it is in the electric field context.

\begin{figure}
\includegraphics[width=3.5in]{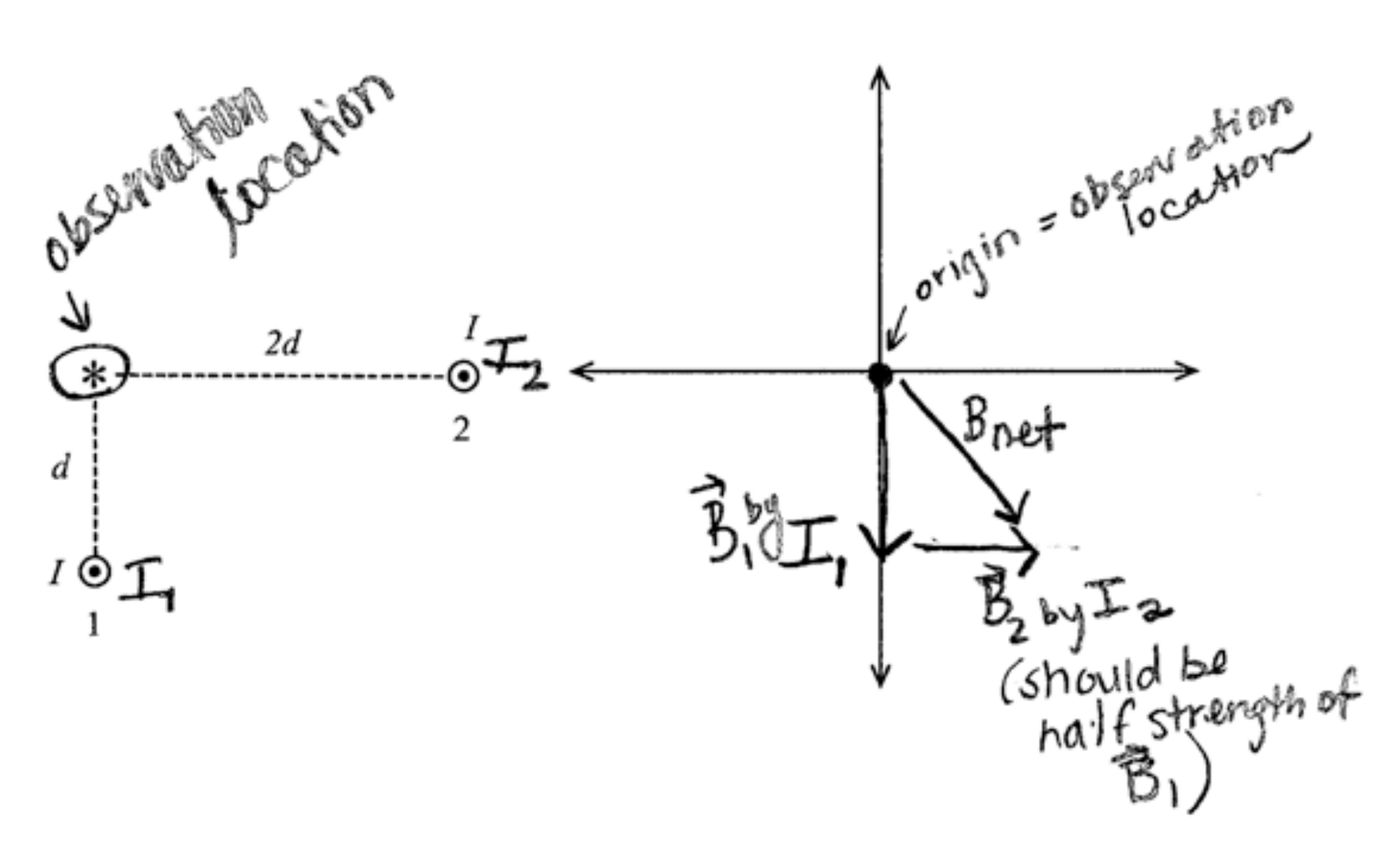}
\caption{Gilma's solution to the magnetic field question (Fig.~\ref{quiz6}). Vector placement was coded as `Asterisk' because the origin of the coordinate system was explicitly identified as the asterisk (observation location). Since the fields from each source point toward the source, this was coded as a  `Direction Error'.}
\label{Gilma}
\end{figure}

\begin{figure}
\includegraphics[width=3.5in]{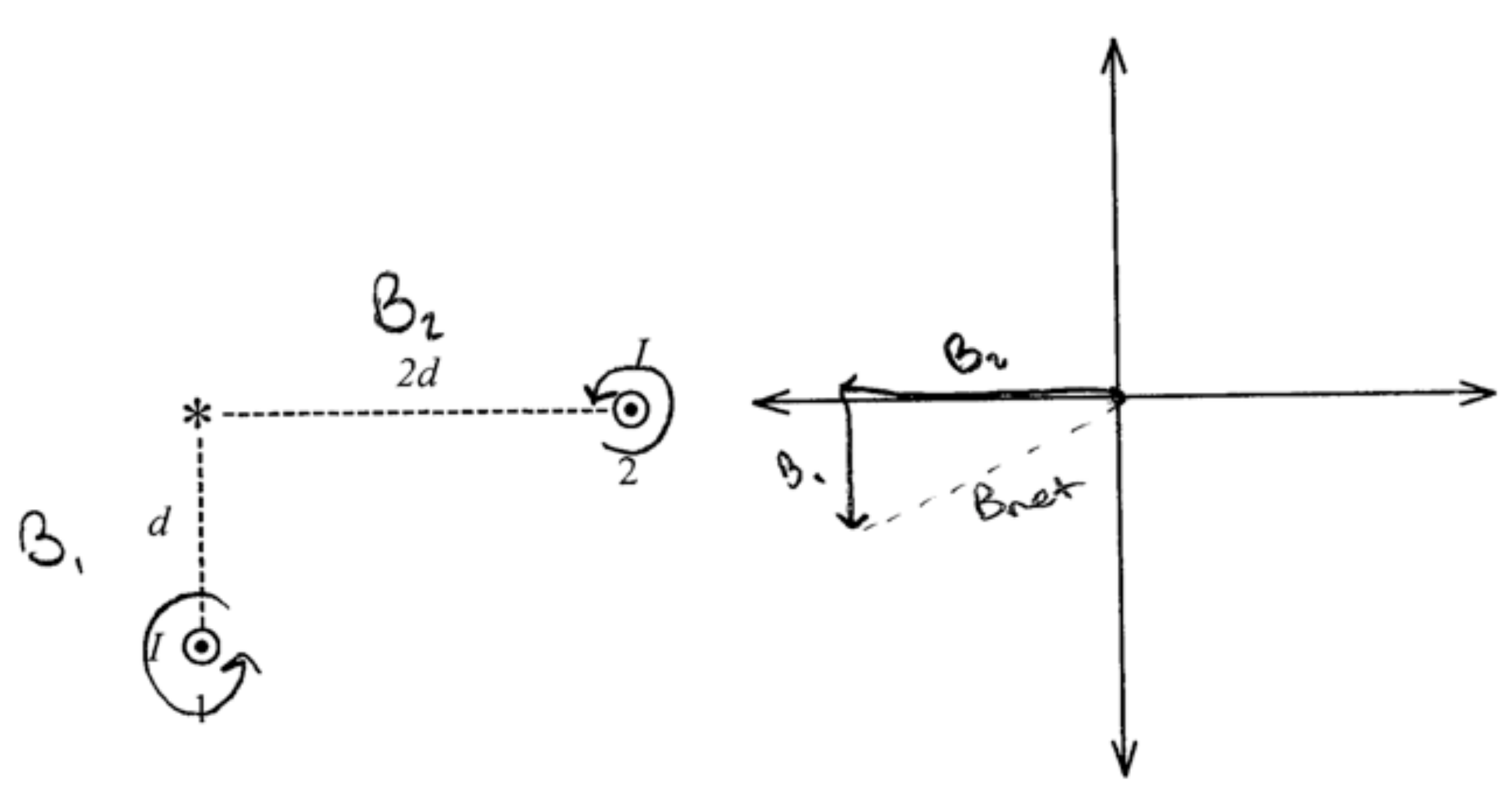}
\caption{Leslie's solution to the magnetic field question (Fig.~\ref{quiz6}). Vector placement was coded as `Without Asterisk' since there is no connection between the separate coordinate system and the asterisk (observation location). This is an example of a student who drew the overall pattern of the magnetic field, but did not interpret it appropriately to find the field at the observation location. Although both the directions and relative magnitude are incorrect, this was coded as a `Direction Error' since direction errors took precedence.}
\label{Leslie}
\end{figure}

Figures~\ref{Gilma} and \ref{Leslie} show typical responses from Quiz 2 coded as `Direction Error'. Note that both ``Gilma'' (Fig.~\ref{Gilma}) and ``Leslie'' (Fig.~\ref{Leslie}) draw vectors that are parallel instead of perpendicular to the separation vector. One interpretation of this is that students are confusing magnetic fields with electric fields, which point toward or away from source charges. This is consistent with research suggesting that student confuse electric and magnetic concepts \cite{Maloney:2001,Scaife:2011,Kustusch:2015}.

The circles drawn around the sources in Leslie's solution (Fig.~\ref{Leslie}) suggest another interpretation, consistent with \citet{Kustusch:2015}, that these direction errors are due to not appropriately connecting a right-hand rule to the observation location. Leslie was one of seventeen students who explicitly drew circles around each current-carrying wire in the provided figure. These circles appear to represent the overall pattern of the magnetic field around each source and for all of these students, the direction of circular pattern (counter-clockwise) was consistent with the correct use of a right-hand rule for these sources. However, despite recognizing the overall pattern of the magnetic field, Leslie does not to correctly identify where along the circular pattern the observation location is and how this would translate to a vector at that location. Six of the seventeen students who drew these circular patterns were coded with a `Direction Error'.

\begin{figure}
\includegraphics[width=3.5in]{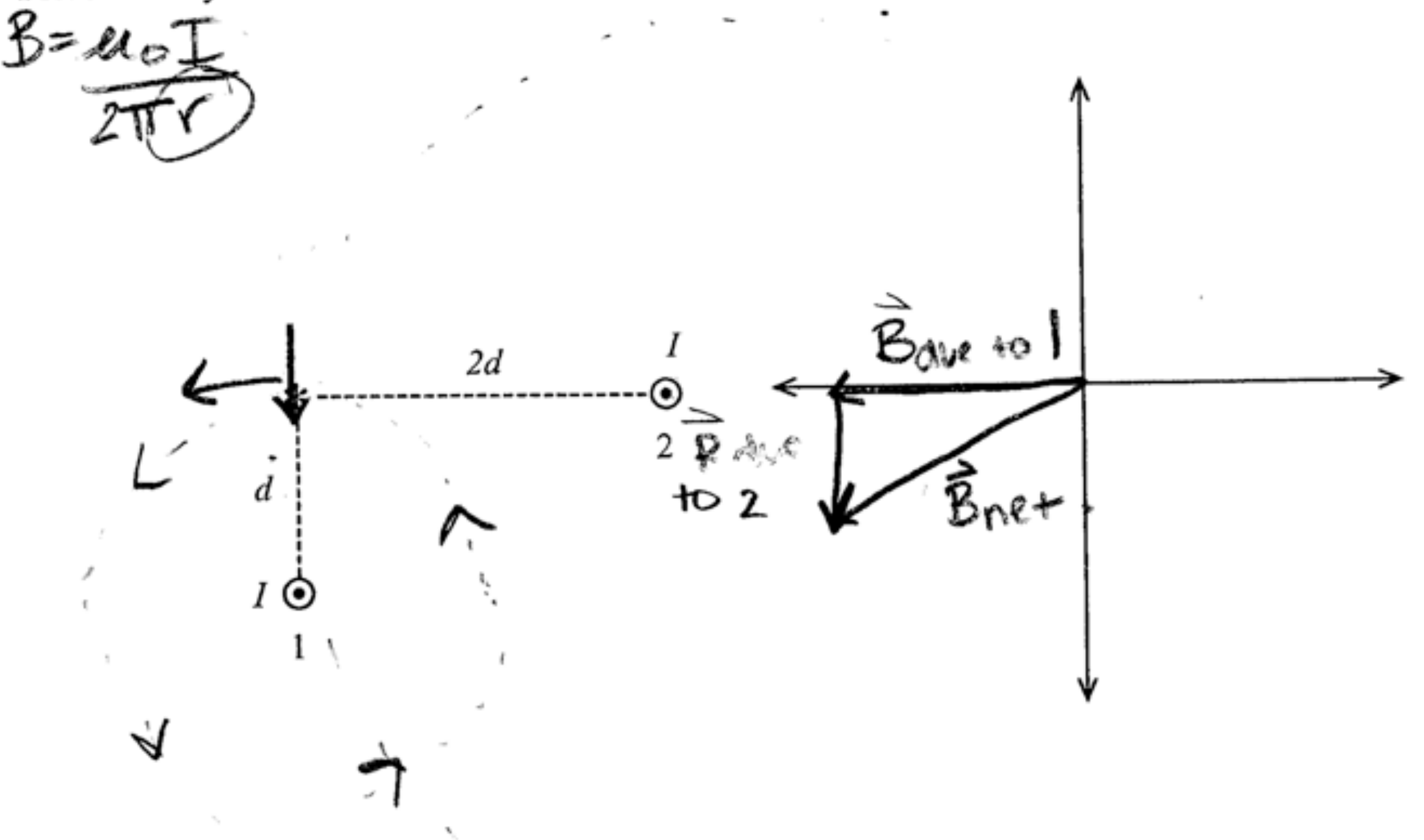}
\caption{Norma's solution to the magnetic field question (Fig.~\ref{quiz6}). Although not shown, she also wrote ``Closer wire has greater magnetic field.'' This is an example of a student who explicitly showed radii of the circle (dotted lines) extending to the observation location.}
\label{Norma}
\end{figure}

In contrast, most of the students (11 out of 17) who drew the overall pattern of the field were able to use this approach productively. For example, ``Norma'' (see Fig.~\ref{Norma}) draws a vector tangent to each circle at the observation. In addition, she uses circles of different radii and wrote $B=\frac{\mu_0I}{2\pi r}$, with the $r$ circled and wrote ``Closer wire has greater magnetic field'' (not shown). She uses this to correctly identify the relative magnitude of the two vectors shown on the separate coordinate system.
Norma represents a student who has a good understanding of observation location and how the measurement of magnetic field depends on the location of the observer.

Finally, we did see errors in relative magnitude as well, although it is difficult to compare the two quizzes since we gave direction errors precedence over magnitude errors and the two quizzes had different functional dependence on distance. Leslie's solution (Fig.~\ref{Leslie}) is an example of one that was coded as a `Direction Error', but also shows an incorrect relative magnitude. She drew vectors that are directly proportional to the distance instead of inversely proportional. There were other `Magnitude Errors' where students would draw the vectors with equal length, indicating that they did not fully recognize that the strength of the field depends on the distance between the source and the observation location.

\section{Conclusions}
In this preliminary look at algebra-based students' understanding of the observation location, we analyzed three aspects of student responses: vector placement, direction, and magnitude. Our results are consistent with previous research, which primarily focused on calculus-based students \cite{Garza:2012,Scaife:2011,Kustusch:2015}.

We find that errors with vector placement are somewhat prevalent in both electric and magnetic field contexts. The evidence is consistent with previous research that students may be confusing field and electric concepts \cite{Garza:2012,Kustusch:2015}. In addition, students may be confusing the role of source and test charges. It is typical to introduce electric field by talking about the force a test charge would feel from source charges. While there are important pedagogical reasons for this choice, some students appear to have difficulty differentiating between these ideas. The concept of an observation location is typically not introduced until magnetic fields (if at all) and more explicit attention to the role of observation location earlier may help students to differentiate better between source and test charges, as well as between field and force. 

Errors with direction, somewhat prevalent on the electric field questions, become very prevalent on the magnetic field question.
These results show that students who can appropriately account for observation location with electric fields may not be able to do so with magnetic field. This is likely due to previously documented interference between electric and magnetic concepts~\cite{Scaife:2011} or to not being able to appropriately interpret the results of a right-hand rule~\cite{Kustusch:2015}.



We intend to use these results to guide our qualitative analysis of student discussions in classroom video.
We plan to explore possible mechanism(s) for the issues presented here, to assess how and to what extent classroom activities elicit and address student difficulties, redesign classroom activities to address student difficulties and document how these activities can be enacted in the classroom. In particular, at least two sections of the class conducted activities where students were asked to use arrows to represent the electric or magnetic field at their location from a source (represented by a ball or PVC pipe) \cite{Kustusch:2014}. We plan to look at the student discussions during these activities to provide further insight into how they understand the spatial nature of electric and magnetic fields.

\begin{acknowledgments}
This project was partially funded by the Undergraduate Research Assistant Program and the Undergraduate Summer Research Program at DePaul University.
\end{acknowledgments}

\bibliography{PERC15_Bryant.bib}

\end{document}